\documentstyle[12pt]{article}
\def\\{\setminus}
\def\({\Bigl(}
\def\){\Bigr)}
\def\|{\Big|}
\def\o{\circ}
\def\x{\times}

\def\ox{\otimes}

\def\pl{\oplus}

\def\mid{\big\bracevert}

\def\then{~\Rightarrow~}

\def\and{\wedge}

\def\m{\bullet}

\def\A{{\,{\rm A\kern-.55emA}}}
\def\C{{\,{\rm I\kern-.55emC}}}
\def\E{{\,{\rm I\kern-.2emE}}}
\def\I{{\,{\rm I\kern-.2emI}}}
\def\K{{\,{\rm I\kern-.2emK}}}
\def\L{{\,{\rm I\kern-.2emL}}}
\def\M{{\,{\rm I\kern-.16emM}}}
\def\N{{\,{\rm I\kern-.16emN}}}
\def\Q{{\,{\rm I\kern-.5emQ}}}
\def\R{{\,{\rm I\kern-.2emR}}}
\def\S{{\,{\rm I\kern-.42emS}}}
\def\T{{\,{\rm I\kern-.37emT}}}
\def\Z{{\,{\rm Z\kern-.35emZ}}}
\def\rin{{\,\in\kern-.42em\in}}

\def\tr{{\,{\rm tr }\,}}

\def\FIX{\,\hbox{FIX}}
\def\STAB{\,\hbox{STAB}}

\def\sup{\supseteq}
\def\supnoteq{\supset}

\def\p{\partial}

\def\al{\alpha}  \def\be{\beta} \def\ga{\gamma}
\def\de{\delta}  \def\ep{\epsilon}  
\def\th{\theta}    \def\io{\iota}
   \def\la{\lambda}   \def\si{\sigma}
   \def\om{\omega} 
\def\phi{\varphi}    
    \def\La{\Lambda}

\let\rvec=\vec        
\def\vec#1{\underline{\bf vec}_{#1}}

\def\d{\displaystyle}
\def\t{\textstyle}
\def\s{\scriptstyle}

\def\Ffo#1{\angle{#1}_{{}_{{\rm F}}}   }  
\def\norm#1{\parallel\!#1\!\parallel}
\def\d#1{\check{#1}}
\def\angle#1{\langle#1\rangle}

\def\lrvec#1{
{\textstyle^{^{^{\leftrightarrow}}}\displaystyle\!\!\!\!\!#1}}
\def\brack#1{\lbrack#1\rbrack}
\def\brace#1{\lbrace#1\rbrace}

\def\ul#1{\underline{#1}}
\def\uul#1{\ul{\ul{#1}}}
\def\ol#1{\overline{#1}}
\def\bl#1{{\bf #1}}
\def\cl#1{{\cal #1}}

\def\ro#1{{\rm #1}}
\def\mini#1{{\scriptstyle#1\displaystyle}}

\def\sprod#1#2{\langle#1|#2\rangle}
\def\com#1#2{\lbrack#1,#2\rbrack}

\def\acom#1#2{\{#1,#2\}}

\def\expv#1#2#3{\langle#1|#2|#3\rangle}

\def\mape{\longmapsto}

\begin{document}

\thispagestyle{empty}
\hfill MPI-PhT/95-32 (april 95)

\begin{center}

{\large
{\Huge Quantum Fields \\ \`{a} la  \\ Sylvester and Witt }

\vspace{3ex}
 H. Saller, R. Breuninger, M. Haft\footnote{
e-mail adresses: \\
H.S.: hns@dmumpiwh and hns@mppmu.mpg.de \\
R.B.: rab@mppmu.mpg.de \\
M.H.: mah@mppmu.mpg.de
} \\[5ex]
{\it Max-Planck-Institut f\"{u}r Physik \\
 -- Werner-Heisenberg-Institut -- \\
 P.O.Box 40 12 12, 80712 Munich (Germany) }
}

\end{center}

\vspace{5ex}
\centerline{Abstract}

A structural explanation of the coupling constants in the standard model, i.e
the fine structure constant and the Weinberg angle, and of the gauge fixing
contributions  is given in terms of symmetries and representation theory.
The coupling constants
are normalizations of Lorentz invariantly embedded little groups
(spin and polarization) arising
in a harmonic analysis of quantum vector fields.
It is shown that
the harmonic analysis of massless fields requires an extension of the
familiar Fourier decomposition, containing also indefinite unitary
nondecomposable time representations. This is illustrated by the
nonprobabilistic contributions in the electromagnetic field.

\newpage

\section{Introduction}

The standard model dynamics  for the electroweak
interactions is quantified by masses and coupling constants -
especially the Fermi mass $\underline{M}$ in the symmetry breakdown mechanism,
the electromagnetic coupling constant (Sommerfeld's
fine structure constant $\al_e$) and its relation to the weak
coupling constants (Weinberg angle $\th_w$). This paper
is an attempt to understand in the standard model
- at least qualitatively
in terms of symmetries and representation theory -
the origin of those structurally and numerically important scales.
Although this structural explanation does not allow to calculate these
parameters within the standard model, we believe that it is a necessary step
towards a more fundamental quantum field theory in which they
may prove to be derivable and therefore calculable.

All relevant constants appear as normalizations
for  the Casimir invariants of the little
groups in the Lorentz group defined by a harmonic analysis.
Due to Wigner\cite{Wig39}, one classifies
relativistic quantum fields with repect to particles by
mass-momenta $(m,\rvec q)$  on causal energy-momenta orbits
$q^2=m^2\ge0$. Two different particle types are possible because of
the indefinite signature $\ro{sign}~g=(1,3)$ of the
Lorentz bilinear form $g$. Particles with strictly positive
mass $m^2>0$ obey an $SO(3)$ spin classification
in the Lorentz group $SO(1,3)$, particles with vanishing
mass $m^2=0$ are classified with respect to a polarization $SO(2)$ only.
These little groups  prove to be the stability groups of
decompositions for Minkowski translations, in the
massive case a Sylvester decomposition into time and space
translations, in the massless case a  Witt decomposition\cite{Bou59}
into two lightlike translations and 2-dimensional space translations.
Therefore massive and massless fields will be called fields \`a la Sylvester
and Witt resp.

The time axis is defined in the translations decompositions. Then
the stability groups are compatible with the representation of the
time translations, classified with respect
to their unitarity properties as follows.

Massive fields with spin have an
irreducible positive unitary $U(1)$ classification
for their time development
as familiar from quantum mechanical bound states,
e.g. the harmonic oscillator. The invariant energies
(frequencies) of the quantum mechanical case
arise as invariant masses in the case of massive relativistic fields.

The situation is completely different
for massless fields with circularity (polarization).
Here, especially in massless vector
fields, e.g. the electromagnetic field,
harmonic components without a
probabilistic particle interpretation occur (interactions without
particle parametrization). New structures
arise, e.g. dipole contributions in the propagator or
'gauge fixing constants' which have to
be nontrivial on the one hand, but which are experimentally irrelevant
on the other hand. This shows that
the time representation structure of relativistic fields
is more general than their classification into particle interpretable parts
('every particle is a field, but not every field is a particle').
Therefore, in extension of Wigners definition of particles as
irreducible positive
unitary representations also nondecomposable
indefinite unitary ones have to be
considered in the time development classification of
relativistic fields\cite{Sal94b}.
We shall give in some detail the connection between
those indefinite $U(1,1)$ structures and the corresponding
harmonic components in the massless fields - e.g. the
Coulomb components. It will become clear in this analysis
that the 'gauge fixing constant' has its
structural origin in the basic dependent and, therefore,
physically irrelevant normalization
of the nilpotent contribution in the indefinite unitary
Hamiltonian.
It is also shown that even in the massless case the harmonic analysis
requires a specification of a rest system and therewith the introduction
of a mass scale.

\section{Quantum  Fields \`{a} la Sylvester}

Relativistic quantum fields \`a la Sylvester are  completely parametrizable
with massive  particles. As a prominent example we discuss the properties of
a massive
vector field, e.g. the $Z$-boson field in the standard model.

A free massive vector field $(\bl Z^j(x))_{\mini{j=0,1,2,3}}$
can be  introduced
with a  Lagrangian leading to its dynamical behaviour
\begin{eqnarray}
&&\cl L(\bl Z^j,\bl G^{jk})=\bl G^{jk}{\p_j \bl Z_k-\p_k \bl Z_j\over2}
+m(\la{\bl G^{jk}\bl G_{jk}\over 4}+{\bl Z^j\bl Z_j\over2\la}) \\
&&\ep^{jk}_{lr}\p^l\bl Z^r=\p^j\bl Z^k-\p^k\bl Z^j=m\la\bl G^{kj},~~
\p_k \bl G^{jk}=-{m\over\la}\bl Z^j ,~~m\la>0
\end{eqnarray}
In this 1st order derivative formalism $(\bl Z,\bl G)$ is a
canonical pair. $\ep^{jk}_{lr}=
\de^j_l\de^k_r -\de^k_l\de^j_r$  are the Clebsch-Gordan coefficients
for the projection of the $D^{({1\over2}|{1\over2})}$ Lorentz vector
representations $\p$ and $\bl Z$ to the $D^{(1|0)}\pl D^{(0|1)}$
Lorentz tensor representation $\bl G$.

By using three 'natural' scales, $\hbar$ (Plancks action unit),
$c$ (highest action velocity) and $\ul M$ -
an unspecified mass scale, e.g. the  symmetry breakdown mass
of the standard model,
all parameters and fields are  dimensionless.

The dynamics involves two characteristic numbers:
The parameter $m$ is the {\it particle mass}. The parameter $\la$
can be normalized away in a free theory
with a dilatation transformation
(the omitted Lorentz indices have to be inserted as above)
\begin{equation}
 \begin{array}{l}
  \mini{{1\over\sqrt{m\la}}}\bl Z=\bl Z' \\
   \mini{\sqrt{m\la}}\bl G=\bl G'
 \end{array}
 \then
 \left\{ \begin{array}{l}
  \cl L(\bl Z,\bl G)=\bl G'{\p \bl Z'\over2}
  +{\bl G'\bl G'\over 4}+m^2{\bl Z'\bl Z'\over2} \\
  \p\bl Z'=\bl G',~~ \p \bl G'=-m^2\bl Z'
 \end{array} \right.
\end{equation}
However in an interaction,
e.g. of gauge type with a Dirac field $\bl\Psi$,
the dilatation  factor arises  as a {\it coupling constant}, e.g.
\begin{equation}
\bl Z_j\ol{\bl \Psi}\ga^j\bl\Psi=
g_Z\bl Z'_j\ol{\bl \Psi}\ga^j\bl\Psi,~~ g_Z^2=m\la
\end{equation}

The dynamical structure of the massive field
should be seen in comparison with a harmonic oscillator
with the canonical position-momentum pair $\bl x(t),\bl p(t)$
\begin{eqnarray}
& L(\bl x,\bl p)
=\bl p{d\bl x\over dt}-({\bl p^2\over2M}+k{\bl x^2\over2})\cr
&{d\bl x\over dt}={\bl p\over M},~~{d\bl p\over dt}=-k\bl x,~~
kM>0
\end{eqnarray}
In the quantum mechanical model  $\hbar$ is used as natural unit
with equal time commutator $\com{i\bl p(t)}{\bl x(t)}=1$.
The mass $M$ and the  spring constant $k$ give
the scales for time $\mini{{1\over\om}}$
and length $\ell$
\begin{eqnarray}
&& \om^2=\mini{{k\over M}},~~
\ell^4=\mini{{1\over kM}} =\mini{{\om^2\over k^2}}\cr
&& L(\bl x,\bl p)
=\bl p{d\bl x\over dt}-\om(\mini{{1\over\sqrt{kM}}}
{\bl p^2\over2}+\mini{\sqrt{kM}}{\bl x^2\over2})
\end{eqnarray}

The frequency decomposition of the harmonic oscillator displays its
unitary time  development  $e^{i\om t}\in U(1)=e^{i\R}$ (phase property)
and its dilatation property ${\om\over k}\in e^\R$
\begin{equation}
\bl x(t)=
\mini{ \sqrt{{\om\over k}} }
{e^{i\om t}\ro u+e^{-i\om t}\ro u^\star\over\sqrt2},~~
-i\bl p(t)=
\mini{ \sqrt{{k\over \om}} }
{e^{i\om t}\ro u-e^{-i\om t}\ro u^\star\over\sqrt2}
\end{equation}
The analogue structure
is reflected in the harmonic analysis of the relativistic field with
$\mini{q_0=\sqrt{m^2+\rvec q^2}}$
\begin{eqnarray}
\bl Z^j(x)&=&\mini{\int {d^3q\over\sqrt{(2\pi)^3q_0}}}
\La(\mini{q,m})^j_a
{}~\mini{\sqrt{m\la}}
{e^{ixq}~ \ro U^a(\rvec q)+e^{-ixq}~\ro U^{\star a}(\rvec q)\over\sqrt2}\cr
i\bl G^{kj}(x)&=&
{}~\mini{\int {d^3q\over\sqrt{(2\pi)^3q_0}}}
\ep_{lr}^{kj}q^l
\La(\mini{q,m})_a^r
\mini{ {1\over \sqrt{m\la }} }
{e^{ixq}~\ro U^a(\rvec q)-e^{-ixq}~\ro U^{\star a}(\rvec q)\over\sqrt2}
\end{eqnarray}
$\La(q,m)$ transforms from the general momenta $q$ with $q^2=m^2$ to the
rest frame defined by the massive field. It
mediates between the regime of the orthochronous
Lorentz group with variables $\bl Z^j(x)$ and the spin regime with
variables $\ro U^a(\rvec q),\ro U^{\star a}(\rvec q)$.
We call $\La(q,m)$ a transmutator. It is a  representative
\footnote{For a group $G$ with subgroup $H$ a coset
$gH\in G/H$ has representatives $x\in gH$, denoted by $x\rin G/H$.}
for a class
of the related real 3-dimensional coset space, completely
parametrizable by three noncompact
momenta $\mini{({q^a\over m})_{a=1,2,3}}$
\begin{equation}
\La(\mini{q,m})^k_{0,a}
\cong\mini{{1\over m}}{\pmatrix{
q^0&\rvec q\cr
\rvec q&\de^{bc}m+{q^bq^c\over q^0+m}\cr}}
\rin SO^+(1,3)/SO(3)
\end{equation}
The Lorentz-orbits for the $SO(3)$-invariant bilinear form $\de^{ab}$
give the spin-1-pro\-jec\-tors $\mini{-\eta^{kj}+{q^kq^j\over m^2}}$
\begin{eqnarray}
&&\La(\mini{q,m})^k_am^2\de^{ab}\La(\mini{ q,m})_b^j
=-m^2\eta^{kj}+q^kq^j,~~-\eta^{kj}\cong{\pmatrix{
-1&0&0&0\cr
0&1&0&0\cr
0&0&1&0\cr
0&0&0&1\cr}}\cr
&&\La(\mini{q,m})^k_0m=q^k ,~~
\La(\mini{q,m})^k_0\eta_{kj}\La(\mini{ q,m})^j_b=0
\end{eqnarray}
The mass $m^2$ of the vector particle turns out to be
the {\it intrinsic scale
for the spin group}
$SO(3)$, i.e. the normalization of its Casimir invariant
$\si^{ab}=m^2\de^{ab}$ in the vector field representation.

The time dependent quantization of the harmonic oscillator
with the shorthand notation $\com a b(t-s)=\com{a(s)}{b(t)}$
\begin{eqnarray}
\com{\ro u^\star}{\ro u}&=&1\cr
\then {  \pmatrix{
\s\com{i\bl p}{\bl x}&\s\com{\bl x}{\bl x}\cr
\s\com{\bl p}{\bl p}&\s\com{\bl x}{-i\bl p}\cr}}(t)&=&
{  \pmatrix{
\s \cos\om t&\s{\om\over k}i\sin\om t\cr
\s{k\over \om}i\sin\om t&\s\cos\om t\cr}}
\ =\ { \pmatrix{
\s-i{d\over\om dt}&\s{\om\over k}\cr
\s{k\over\om} &\s-i{d\over\om dt}\cr}}i\sin\om t
\end{eqnarray}
arises for the vector field as follows
\begin{eqnarray}
\com{\ro U^{\star a}(\rvec q)}{\ro U^b(\rvec p)}
&=&\de^{ab}\de(\rvec p-\rvec q)\cr
\then { \pmatrix{
\s\com{i\bl G^{kl}}{\bl Z^j}&
\s\com {\bl Z^k}{\bl Z^j}\cr
\s\com{\bl G^{kl}}{\bl G^{jn}}&\s\com {\bl Z^k}{-i\bl G^{jn}}\cr}}(x)
&=&{ \pmatrix{
\s-i\ep_{ut}^{lk}\de^j_s\p^u &\s m\la\de_t^k\de_s^j\cr
\s-{1\over \la}\ep_{ut}^{lk}\ep_{rs}^{nj}{\p^r\p^u\over m}
&\s-i\de^k_t\ep_{rs}^{nj}\p^r\cr}}
\t {\com {\bl Z^t}{\bl Z^s}(x)\over m\la}
\end{eqnarray}
with the vector field  quantization as the analogue  to
$\com{\bl x}{\bl x}(t)=\mini{{\om\over k}}i\sin\om t$ given by
\begin{eqnarray}
\com {\bl Z^k}{\bl Z^j} (x)
&=&\mini{\int {d^4 q\over(2\pi)^3}e^{ixq}
m\la (-\eta^{kj}+{q^kq^j\over m^2})}
\ep(q^0)\de(q^2-m^2)
 \cr
&=&\mini{\int {d^3 qe^{-i\rvec x\rvec q}\over(2\pi)^3q^0}}
\La(q,m)_a^k~\brack{\bl{ZZ}}^{ab}(x_0)~\La(q,m)^j_b\cr
\brack{\bl{ZZ}}^{ab}(x_0)&=&m\la\de^{ab}i\sin q^0x_0
\end{eqnarray}

The Fock expectation values, abbreviated  by
$\Ffo{\acom ab}(t-s)=\expv 0 {\acom{a(s)}{b(t)}}0$,
for the creation operator $\ro u$ and its annihilation
partner $\ro u^\star$
\begin{eqnarray}
\expv 0{\ro u^\star \ro u}0&=&1\cr
\then{ \pmatrix{
\s\Ffo{\acom{i\bl p}{\bl x}}&\s\Ffo{\acom{\bl x}{\bl x}}\cr
\s\Ffo{\acom{\bl p}{\bl p}}&\s\Ffo{\acom{\bl x}{-i\bl p}}\cr}}(t)
&=&{ \pmatrix{
\s i\sin\om t&\s{\om\over k}\cos\om t\cr
\s{k\over\om} \cos\om t&\s i\sin\om t\cr}}
\ =\ { \pmatrix{
\s-i{d\over\om dt}&\s{\om\over k}\cr
\s{k\over\om}&\s-i{d\over\om dt}\cr}}\cos\om t
\end{eqnarray}
read for the particle field
\begin{eqnarray}
\expv 0{\ro U^{\star a}(\rvec q)\ro U^b(\rvec p)} 0
&=&\de^{ab}\de(\rvec p-\rvec q)\cr
\then { \pmatrix{
\s\Ffo{\acom{i\bl G^{kl}}{\bl Z^j}}&\s\Ffo{\acom {\bl Z^k}{\bl Z^j}}\cr
\s\Ffo{\acom{\bl G^{kl}}{\bl G^{jn}}}&
\s\Ffo{\acom {\bl Z^k}{-i\bl G^{jn}}}\cr}}(x)
&=&{ \pmatrix{
\s-i\ep_{ut}^{lk}\de^j_s\p^u   & \s m\la\de_t^k\de_s^j\cr
\s-{1\over \la}\ep_{ut}^{lk}\ep_{rs}^{nj}{\p^r\p^u\over m}
&\s-i\de^k_t\ep_{rs}^{nj}\p^r\cr}}
\t{\Ffo{\acom {\bl Z^t}{\bl Z^s}}(x)\over m\la}
\end{eqnarray}
The relativistic analogue of
$\Ffo{\acom{\bl x}{\bl x}}(t)=\mini{{\om\over k}}\cos\om t$
is given by
\begin{eqnarray}
\Ffo{\acom {\bl Z^k}{\bl Z^j}}(x)
&=&\mini{\int {d^4 q\over(2\pi)^3}e^{ixq}
m\la (-\eta^{kj}+{q^kq^j\over m^2}) }\de(q^2-m^2)           \cr
&=&\mini{\int {d^3 qe^{-i\rvec x\rvec q}\over(2\pi)^3q^0}}
\La(q,m)_a^k~\Ffo{\brace{\bl{ZZ}}^{ab}}(x_0)~\La(q,m)^j_b\cr
\Ffo{\brace{\bl{ZZ}}^{ab}}(x_0)&=&m\la\de^{ab}\cos q^0x_0
\end{eqnarray}

The intrinsic length scale of the oscillator
\begin{equation}
\Ffo{\acom{\bl x}{\bl x}}(0)=2\expv 0{\bl x^2}0
=2\norm{\bl x}^2={\om\over k}=\ell^2
\end{equation}
has its analogue in the  coupling constant $g_Z^2=m\la$.
It is suggested to identify
the square of the massive vector field  is identified
with the $SO(3)$-Casimir invariant. Then the coupling constant
coincides with the mass
\begin{equation}
\Ffo{\brace{\bl{ZZ}}^{ab}}(0)=\si^{ab}=m^2\de^{ab}
\then \la=m,~g_Z^2=m^2
\end{equation}

\section{Decompositions \`a la Sylvester and Witt}

Relativistic fields map space-time translations $\M\cong\R^4$
into vectors of complex spaces $V\cong\C^n$.
Being complex valued is essential for
the quantum structure of fields
(phases, probability 'amplitudes', scalar products
etc.). Relativistic fields are   compatible with the
action of the  orthochronous Lorentz group $SL(\C^2)$
on the Minkowski space $\M$ -
as phaseless real group $SO^+(1,3)\cong SL(\C^2)/\{\pm\bl 1_2\}$
and on the value space  $V$ for the field, e.g. for a vector field
$\La^i_j\bl Z^j(x)=\bl Z^i(\La(x))$ with
$\La=D^{({1\over2}|{1\over2})}(\la)$.

The harmonic analysis classifies  a relativistic field with respect to
unitary  representations of time translations,
valued in the general linear  group $GL(V)$. In extension of
Wigners definition of particles with positive unitary representations
also indefinite unitary ones will be cosidered.
A harmonic analysis relies  on finite
dimensional  representations of a
direct product subgroup $G_{\ro{stab}}\x\T$
of the whole semidirect Poincar\'e group $SO^+(1,3)\x_s\M$
with only the time translations
$\T\cong\R$. The  subgroup
$G_{\ro{stab}}$ of the homogeneous Lorentz group proves to be
the stability group
\footnote{
The fixgroup in a group $G$ acting on a set  $S$
keeps every element of the set invariant
$\FIX_SG=\{g\in G\mid g\m x=x\hbox{ for all }x\in S\}$,
the stability group only the whole set
$\STAB_SG=\{g\in G\mid g\m S=S\}\sup \FIX_SG$.}
of a decomposition of the
Minkowski translations, determined by the
'masses' $m^2>0,=0,<0$  of the relativistic fields.

For the causal case with masses $m^2\ge0$  -
then only real energies $\mini{q^0=\sqrt{m^2+\rvec q^2}}$ occur -
the space-time translations can be decomposed
either \`a la Sylvester
\footnote{James Joseph Sylvester (1814-1897),
Ernst Witt (1911-...)}
for $m^2>0$ or \`a la Witt for $m^2=0$.

{}From the momentum orbit $q^2=m^2$ of a
massive field
one can distinguish one timelike vector $\ul{\bl e}_0$
\begin{equation}
q^2=m^2>0\then\ul q= \ul{\bl e}_0\cong (m,0,0,0)
\end{equation}
This leads to the definition of
a rest system
and therewith a {\it Sylvester decomposition}\cite{Bou59}
$\M\cong\T\pl\S$ into time and space translations.
The fixgroup of the distinguished time translation  $\ul{\bl e}_0$ is also the
stability group of the decomposition with $\T=\R\ul{\bl e}_0$, it is the
compact rotation group (spin group)
\begin{eqnarray}
\FIX_\T SO^+(1,3)&=&\STAB_{\T\pl\S}SO^+(1,3)= SO(3)\cong
 SU(2)/\{\pm\bl 1_2\}\cr
G_{\ro{stab}} \x \T&\cong& SO(3)\x \R
\end{eqnarray}

{}From the momentum orbit $q^2=0$ of a
massless field
one can distinguish one lightlike vector $\ul{\bl e}_+$
\begin{equation}
q^2=0,~q\ne0  \then
\ul q=\ul{\bl e}_+\cong  {\mu\over\sqrt2}(1,0,0, 1)
\end{equation}
The mass parameter $\mu^2>0$ is not given with the massless orbit.
The fixgroup of one lightlike translation, e.g. $\ul{\bl e}_+$,
and therewith of the 1-dimensional subspace $\L_+=\R\ul{\bl e}_+$
is the noncompact semidirect Euklidian group in two real dimensions
\begin{equation}
\FIX_{\L_+} SO^+(1,3) =SO(2)\x_s\R^2
\end{equation}
This fixgroup is no stability group for a vector space decomposition,
since the direct complement $\L_-\pl\S^2$ is not invariant.

Since a  time axis cannot be distinguished by using
only $\ul{\bl e}_+$,
one has to consider for massless fields
the fixgroup of  two independent lightlike translations, e.g. $\ul{\bl e}_\pm$.
It is the
fixgroup af all lightlike translations $\L=\L_+\pl\L_-\cong\R^2 $ and therewith
the stability group of a {\it Witt
decomposition}\cite{Bou59} into three direct summands
$\M\cong\L_+\pl\L_-\pl\S^2$ with  2-dimensional space translations $\S^2$
and is given by the compact axial group (polarization or circularity group)
\begin{eqnarray}
\FIX_\L SO^+(1,3)
&=&\STAB_{\L_+\pl\L_-\pl\S^2 } SO^+(1,3)=SO(2)\cong U(1)\cr
G_{\ro{stab}}\x \T&\cong& SO(2)\x \R
\end{eqnarray}

The definition of $\L$ needs two orbits, e.g. a timelike and a lightlike
one
\begin{eqnarray}
p^2=\ul M^2>0\then \ul p=\ul{\bl e}_0&\cong&(\ul M,0,0,0)\cr
q^2=0,~q\ne0 \then \ul q=\ul{\bl e}_+&\cong&  {\mu\over\sqrt2}(1,0,0, 1) \cr
\then {\mu\over\ul M}\sqrt2 \ul p-\ul q=\ul{\bl e}_-
&\cong&{\mu\over\sqrt2}(1,0,0,-1)
\end{eqnarray}
Neither the mass parameter $\ul M$ nor the mass parameter
$\mu$ arise in the lightlike orbit $q^2=0$.

A Witt decomposition is a subdecomposition of a Sylvester space-time
decomposition as seen in the Lorentz-Sylvester-Witt chain
$SO^+(1,3)\supnoteq SO(3)\supnoteq SO(2)$.

\section{Quantum Spinor Fields \`a la Witt}

In the case of  a Sylvester decomposition of the space time translations
with stability group $SO(3)$,
the real 3-dimensional Sylvester manifold
$SO^+(1,3)/SO(3)\cong$ $SL(\C^2)/SU(2)$
describes the transition to any  Sylvester decomposition.
For a harmonic analysis of  massive fields one starts
from a complex 2-dimensional
Weyl representation $s(q,m)$ of the coset representatives (boosts),
parametrized  by the spacelike momenta $\mini{{\rvec q\over m}}$
\begin{eqnarray}
&&s(\mini{q,m})\o  \bl 1_2\o s^\star(\mini{q,m})={\rho_kq^k\over m}=
\mini{{1\over m}}{ \pmatrix{q^0+q^3& q^1-iq^2\cr
 q^1+iq^2&q^0-q^3\cr}}
,~~\rho_k=(\bl 1_2,\rvec \si),~
\d\rho_k=(\bl 1_2,-\rvec \si)\cr
&&\then \left\{ \begin{array}{l}
s(\mini{q,m})
=s(\mini{q,m})^\star
=\mini{\sqrt{q^0+m\over 2m}\brack{\bl 1_2+{\rvec \si\rvec q\over q^0+m}}} \\
=\mini{{1\over\sqrt{2m(q^0+m)}}}{ \pmatrix{
q^0+m+q^3& q^1-iq^2\cr q^1+iq^2& q^0+m-q^3\cr}}
\rin SL(\C^2)/SU(2),~~q^2=m^2 \end{array}\right.
\end{eqnarray}
The {\it Lorentz-Sylvester transmutators}
in the harmonic analysis of massive fields
are representations of $s(\mini{q,m})$ as used in chapter 1.
for the massive vector field in the vector representation
$\La(q,m)=D^{({1\over2}|{1\over2})}(s(\mini{q,m}))$.

Massless spinor
fields with a Witt decomposition of the space-time translations
involve two  helicity
projectors which can be obtained by the limit $m\to 0$
from the Weyl transmutators of the Sylvester decompositions
\begin{eqnarray}
p_+(q)= p_+(q)^\star&=&
\lim_{m\to0}\mini{\sqrt{{ m\over 2q_0}}} s(\mini{q,m})
=\mini{{1\over2}\brack{\bl 1_2+{\rvec \si\rvec q\over q^0}}}\cr
p_-(q)=p_-(q)^\star&=&
\lim_{m\to0}\mini{\sqrt{{m\over2q_0}}}  s(\mini{q,m})^{-1}
=\mini{{1\over2}\brack{\bl 1_2-{\rvec \si\rvec q\over q^0}}},~~q^2=0
\end{eqnarray}
with two 'compact' momentum parameters $\mini{\rvec q\over q_0}$.

A prominent  spinor Witt field is a massless neutrino field
$(\bl l^A(x))_{\mini{A=1,2}}$ with a left handed Weyl representation
$D^{({1\over2}|0)}$
and Lagrangian
\begin{eqnarray}
&&\cl L(\bl l)={i\over2}\bl l^\star\d\rho_j\lrvec{\p}^j \bl l,~~
i\d\rho_j\p^j\bl l=0
\end{eqnarray}
The harmonic analysis  with particle and antiparticle
creation operators $\ro U(\rvec q)$, $\ro A(\rvec q)$
and particle and antiparticle
annihilation operators $\ro U^\star(\rvec q)$, $\ro A^\star(\rvec q)$
looks like
\begin{eqnarray}
\bl l^{A}(x)
&=&\mini{\int {d^3q\over(2\pi)^{3/2}}}
p_+(q)_C^{A}
\brack{e^{i x q}\ro U^C(\rvec q)+e^{-i xq}\ro A^{\star C}(\rvec q)}
\cr
\bl l^\star_{\dot A}(x)
&=&\mini{\int {d^3q\over(2\pi)^{3/2}} }
p_+(q)_{\dot A}^C~
\brack{e^{-i xq}\ro U^\star_C(\rvec q)+e^{i xq}\ro A_C(\rvec q)}
\end{eqnarray}
We shall show in chapter 4. that this Fourier decomposition is
not quite complete.

One has the anticommutators for the Weyl fields
\begin{eqnarray}
&&\acom{\ro U^\star_A(\rvec q)}{\ro U^B(\rvec p)}
=\de_A^B \de(\rvec p-\rvec q)=
\acom{\ro A^{\star B}(\rvec q)}{\ro A_A(\rvec p)}\cr
\then
&&\acom{\bl l^\star_{\dot B}}{\bl l^{ A}} (x)
=\mini{\rho^j{}_ {\dot B}^{ A}
\int {d^4q\over(2\pi)^3}e^{ixq}q_j\ep(q^0)\de(q^2)}
=\mini{-i\rho^j{}_ {\dot B}^{ A} \p_j
\int {d^3qe^{-i\rvec x\rvec q}\over(2\pi)^3q_0}i\sin x_0q_0}
\end{eqnarray}
and commutator expectation values in the Fock form
\begin{eqnarray}
&& \Ffo{\com{\ro U^\star_A(\rvec q)}{\ro U^B(\rvec p)}}
=\de_A^B \de(\rvec p-\rvec q)
=\Ffo{\com{\ro A^{\star B}(\rvec q)}{\ro A_A(\rvec p)}}\cr
\then
&& \Ffo{\com{\bl l^\star_{\dot B}}{\bl l^{ A}}} (x)
=\mini{\rho^j{}_ {\dot B}^{ A}
\int {d^4q\over(2\pi)^3}e^{ixq}q_j\de(q^2)}
=\mini{-i\rho^j{}_ {\dot B}^{ A} \p_j
\int {d^3qe^{-i\rvec x\rvec q}\over(2\pi)^3q_0}\cos x_0q_0}
\end{eqnarray}

\section{Quantum Vector Fields \`a la Witt}

Relativistic  quantum vector fields \`a la
Witt may contain contributions  without
particle interpretation\cite{Sal94b}.
The most familiar Witt vector field is the
massless electromagnetic field with both  the nonparticle like Coulomb
interaction and the polarized photons, reflecting the
lightlike and spacelike subspaces $\L=\L_+\pl\L_-$
and $\S^2$ resp. in the Witt decomposition of the Minkowski translatione
$\M\cong\L_+\pl\L_-\pl\S^2$.

\subsection{The Field Equations}

For a free massless quantum vector field $\bl A^j(x)$ one has as
Lagrangian and field equations
\begin{eqnarray}
&&\cl L(\bl A^j,\bl F^{jk},\bl L)=
\bl L\p_j\bl A^j+\bl F^{jk}{\p_j\bl A_k-\p_k\bl A_j\over2}
+\mu^2 {\bl F^{jk}\bl F_{jk}\over4}+ \ep\si^2{\bl L^2\over2}\cr
&&\ep^{jk}_{lr}\p^l\bl A^r=\p^j\bl A^k-\p^k\bl A^j=\mu^2 \bl F^{kj},~~
\p_j\bl A^j= -\ep\si^2 \bl L \cr
&&\p_j\bl F^{kj}-\p^k\bl L=0 ,~~\mu^2,\si^2 >0,~ \ep=\pm1\cr
\end{eqnarray}
The canonical partners of the Lorentz vector $\bl A$
are a Lorentz tensor $\bl F$ and
a Lorentz scalar field $\bl L$, often called 'gauge fixing' field.

Again, all fields and parameters are dimensionless
with three universal scales $\hbar,c$ and $\ul M$.
The nontrivial, but otherwise arbitrary
constant $\ep\si^2 $ is usually called the {\it 'gauge fixing'
parameter},
its value is experimentally irrelevant in a 'gauge invariant' dynamics.
The parameter $\mu^2 $ can be normalized away
in a free theory
\begin{equation}
  \begin{array}{l}
   \mini{{1\over\mu}}\bl A=\bl A'\\
   \mini{\mu}\bl F=\bl F',~
   \mini{\mu}\bl L=\bl L'
  \end{array}
  \then
  \left\{ \begin{array}{l}
   \cl L(\bl A,\bl F,\bl L)=
   \bl L'\p\bl A'+\bl F'{\p\bl A'\over2}
   +{\bl F'\bl F'\over4}+ {\ep\si^2\over\mu^2} {\bl L'\bl L'\over2} \\
   \p\bl A'=\bl F',~~
   \p\bl A'=-{\ep\si^2\over\mu^2 }\bl L' \\
   \p\bl F'-\p\bl L'=0
  \end{array} \right.
\end{equation}
Again however in an interaction, e.g. of gauge type, the parameter
$\mu^2 $ arises  as experimentally relevant
{\it gauge coupling constant}
\begin{equation}\bl A_j\ol{\bl \Psi}\ga^j\bl \Psi=
g_A\bl A'_j\ol{\bl \Psi}\ga^j\bl \Psi,~~g_A^2=\mu^2 \end{equation}

\subsection{The quantum mechanical analogue}
There is a  quantum mechanical analogue\cite{Sal89}\cite{Sal92a}
to the  non particle
contributions ('gau\-ge' and Cou\-lomb de\-gree of freedom) in the
time development of the massless vector field using
two positions $(\bl x,\bl x')$
and as canonical partners two momenta $(\bl p,\bl p')$.
The appropriate Lagrangian
\begin{eqnarray}
L(\bl x,\bl x',\bl p,\bl p')
&=&\bl p{d\bl x\over dt} +\bl p'{d\bl x'\over dt}
-\brack{  {\bl p^2\over2 \mini M}+{\bl p'{}^2\over2\mini{M'}}
+k({\bl x\bl p'\over\mini{ M'}}-{\bl x'\bl p\over\mini M}) }\cr
{d\bl x\over dt}&=&-k{\bl x'\over\mini M}+{\bl p\over\mini M},~~ ~~
{d\bl p\over dt}=-k{\bl p'\over\mini{M'}}\cr
{d\bl x'\over dt}&=&k{\bl x\over\mini{M'}}+{\bl p'\over\mini{M'}},~~~~
{d\bl p'\over dt}=k{\bl p\over\mini M},~~~~M M'>0,~k\in\R
\end{eqnarray}
contains two masses $M,M'$
and a frequency $\om=\mini{{k\over\sqrt{MM'}}}$, e.g. derivable
from the equation of motion for the momenta
$\mini{ ({d^2\over dt^2}+\om^2)\bl p=0}$.
$\hbar$ is assumed as universal action scale.

For vanishing frequency $\om=0$, the dynamics is decomposable
into the motion of two free mass points $\mini{ L=\bl p{d\bl x\over dt}
-{\bl p^2\over 2M}}$.

The time dependent quantization reads for the
positions and momenta
\begin{eqnarray}
\!\!\!\!\!\!\!\!\!
\mini{ { \pmatrix{
\s\com{i\bl p}{\bl x}&\s\com{\bl p'}{\bl x}&
\s\com{-i\bl x'}{\bl x}&\s\com{\bl x}{\bl x}\cr
\s\com{-\bl p}{\bl x'}&\s\com{\bl p'}{i\bl x'}&
\s\com{\bl x'}{\bl x'}&\s\com{\bl x}{i\bl x'}\cr
\s\com{i\bl p}{\bl p'}&\s\com{\bl p'}{\bl p'}&
\s\com{-i\bl x'}{\bl p'}&\s\com{\bl x}{\bl p'}\cr
\s\com{\bl p}{\bl p}&\s\com{\bl p'}{-i\bl p}&
\s\com{\bl x'}{-\bl p}&\s\com{\bl x}{-i\bl p}\cr}}(t) }
&=&\mini{ \pmatrix{
\s\cos\om t &\s i\sqrt{{M'\over M}}\sin\om t
&\s-{t\over\sqrt{MM'}}\sin\om t
&\s{it\over M}\cos\om t\cr
\s i\sqrt{{M\over M'}}\sin\om t &\s\cos\om t
&\s{it\over M'}\cos\om t
&\s -{t\over\sqrt{MM'}}\sin\om t\cr
\s 0          &\s 0        &\s  \cos\om t
&\s i\sqrt{{M'\over M}}\sin\om t\cr
\s 0          &\s 0        &\s i\sqrt{{M\over M'}}\sin\om t
&\s \cos\om t\cr}} \cr
&&~~\cr
&=&\mini{
{ \pmatrix{1&{1\over\sqrt{MM'}}{d\over d\om}\cr0&1\cr}}\ox
{ \pmatrix{
\s\cos\om t &\s i\sqrt{{M'\over M}}\sin\om t\cr
\s i\sqrt{{M\over M'}}\sin\om t &\s\cos\om t\cr}}}
\end{eqnarray}

The harmonic analysis is  more complicated as for the
compact $U(1)$-dynamics since one has
a complex 2-dimensional, reducible, but nondecomposable
time representation  in an {\it indefinite} unitary group\cite{Bre95}
${ \pmatrix{1&{it\over M_0}\cr 0&1\cr}}e^{it\om }\in U(1,1)$.
The representations involves both
time eigenvectors $\ro g,\ro g^\x$ ('good')
and  time  nilvectors $\ro b,\ro b^\x$ ('bad')
\begin{eqnarray}
&&{ \pmatrix{\ro b\cr\ro  g\cr}}\mini{(t)}
= e^{it\om}{ \pmatrix{1&{it\over M_0}\cr 0&1\cr}}
{ \pmatrix{\ro b\cr\ro g\cr}},~~
\mini{(\ro g^\x,\ro  b^\x)(t)}
=\mini{(\ro g^\x,\ro  b^\x)}
 e^{-it\om}{ \pmatrix{1&-{it\over M_0}\cr 0&1\cr}} \cr
&~~\cr
&&{ \pmatrix{
\com{\ro g^\x}{\ro b} &\com{\ro b^\x}{\ro b}\cr
\com{\ro g^\x}{\ro g} &\com{\ro b^\x}{\ro g}\cr }}(t)
={ \pmatrix{1&{it\over M_0}\cr 0&1\cr}}e^{it\om }
={ \pmatrix{1&{1\over M_0}{d\over d\om}\cr 0&1\cr}}e^{it\om }
\end{eqnarray}

The Hamiltonian consists of a semisimple part with invariant and therewith
physically relevant frequency $\om$
and a nilpotent part where the mass $M_0$ is
a basis dependent constant, physically irrelevant (unobservable)
\begin{eqnarray}
&& H=\om(
\mini{\sqrt{{M\over M'}}}\bl x\bl p'-\mini{\sqrt{{M'\over M}}}\bl x'\bl p)
+\mini{{1\over M_0}}(
\mini{\sqrt{{M'\over M}}}{\bl p^2\over2}
+\mini{\sqrt{{M'\over M}}}{\bl p^2\over2})\cr
&&\om=\mini{{k\over\sqrt{MM'}}},~~M_0=\ep(M)\sqrt{MM'}\ne0\cr
&&H=\om(\ro b\ro g^\x+\ro g\ro b^\x)+{\ro g\ro g^\x\over M_0}
\end{eqnarray}
The two representations of the Hamiltonian can be transformed into each other
by using  the indefinite unitary harmonic expansion
\begin{eqnarray}
&&\bl x(t)=
\mini{\sqrt{M_0\over M}}
{\ro b(t)+\ro b^\x(t)\over\sqrt2},~~
-i\bl p(t)=
\mini{\sqrt{ M\over M_0}}
{\ro g(t)-\ro g^\x(t)\over \sqrt2}\cr
&&i\bl x'(t)=
\mini{\sqrt{M_0\over M'}}
{\ro b(t)-\ro b^\x(t)\over \sqrt 2},~~
\bl p'(t)=
\mini{\sqrt{ M'\over M_0}}
 {\ro g(t)+\ro g^\x(t)\over\sqrt2}
\end{eqnarray}

The corresponding nontrivial massless field commutators are
\begin{eqnarray}
&&{ \pmatrix{
\com{i\bl F^{kl}}{\bl A^j}&\com {\bl A^k}{\bl A^j}\cr
\com{\bl F^{kl}}{\bl F^{jn}}&\com {\bl A^k}{-i\bl L}\cr}}(x)
={ \pmatrix{
-i\ep_{ut}^{lk}\de^j_s{\p^u\over\mu^2 } &\de_t^k\de_s^j\cr
-\ep_{ut}^{lk}\ep_{rs}^{nj}{\p^r\p^u\over\mu^4}
&i\de^k_t{\p_s\over \ep\si^2 }\cr}}   \com{\bl A^t}{\bl A^s}(x)
\end{eqnarray}
The basic field commutator contains in addition to the
pole structure $\de(q^2)$, relevant for the $U(1)$-time
development (photons), the   characteristic indefinite
dipole structure\cite{Nak90}
$\de'(q^2)$, related to the $U(1,1)$-time development
as will be shown later in more detail
\begin{eqnarray}
\com{\bl A^k}{\bl A^j}(x)&=&
\mini{\int{d^4q\over(2\pi)^3} e^{ixq}\ep(q_0)
\brack{-\mu^2 \eta^{kj}\de(q^2)-(\mu^2 +\ep\si^2 )q^kq^j\de'(q^2)}}\cr
&=&\mini{
\int {d^3 q\over(2\pi)^3q_0}     \brack{
-\mu^2 \eta^{kj}i\sin x_0q_0
+(\mu^2 +\ep\si^2 )\p^k\p^j
i{x_0q_0\cos x_0q_0-\sin x_0q_0\over2q_0^2}   }
e^{-i\rvec x\rvec q}        }  \cr
\com{i\bl F^{kl}}{\bl A_j}(x)&=&
\mini{\int{d^4q\over(2\pi)^3}e^{ixq}\ep^{kl}_{nj}q^n \ep(q_0)\de(q^2)},~~
\com{\bl A^k}{i\bl L}(x)=
\mini{\int{d^4q\over(2\pi)^3} e^{ixq}q^k\ep(q_0)\de(q^2)}
\end{eqnarray}
To perform the energy $q_0$-integration, i.e. to decompose
$q^2=q_0^2-\rvec q^2$,
there has to exist
a time-space decomposition using a rest system.
The terms proportional to $x_0q_0e^{ix_0q_0}$ are characteristic for
the noncompact time representations
involved.
The relation between
compact and  noncompact time representations on the one hand and
distributions, dipoles etc. on the other hand
is obvious in the mechanical model
\begin{eqnarray}
e^{itE}=\int dq_0e^{itq_0}\de(q_0-E),~~&&\de(q_0-E)=\ro{Re}
{i\over\pi}{1\over E+i\ep-q_0}\cr
-ite^{itE}=\int dq_0e^{itq_0}\de'(q_0-E),~~&&\de'(q_0-E)=\ro{Re}
{i\over\pi}{1\over (E+i\ep-q_0)^2}
\end{eqnarray}

The 'gauge fixing' parameter in the combination
$\mu^2+\ep\si^2$ will turn out to be
the analogue to the inverse mass scale
$\mini{{1\over M_0}}$ in the mechanical model.

\subsection{Lorentz-Witt-Transmutation}

The Lorentz-Witt transmutation is performed in two steps. First,
a Lorentz-Sylvester transmutation $s(\mini{p,\ul M})\rin SL(\C^2)/SU(2)$
or $\La(\mini{p,\ul M})\in SO^+(1,3)/SO(3)$ with $p^2=\ul M^2>0$
fixes a time axis (laboratory  system) and a reference mass $\ul M$.
For clarity we denote
\begin{equation}
 \hbox{vector indices for the }
 \left\{ \begin{array}{l}
  \hbox{$SO^+(1,3)$-Lorentz regime: }j=(0,1,2,3) \\
  \hbox{$SO(3)$-Sylvester regime: }\ul j=(0;a)=(0;1,2,3)
 \end{array} \right.
\end{equation}

In addition, one has to use a Sylvester-Witt transmutator
$u(\mini q)$ as element of the  manifold
$SU(2)/U(1)$ which - in the vector representation - rotates
the two distinguished lightlike vectors
$\ul{\bl e}_\pm \cong \mini{{1\over\sqrt2}(1,0,0,\pm 1)}$
into two general lightlike  vectors with a general space direction
$\mini{{\rvec q\over q_0}}$
\begin{eqnarray}
&&u(\mini q) \o{\bl 1_2\pm\si^3\over  2}\o  u^\star(\mini q)=
{\rho_{\ul j} q^{\ul j}_\pm\over q_0}
=\mini{{1\over q^0}}
{ \pmatrix{
q^0\pm q^3&\pm (q^1-iq^2)\cr\pm(q^1+iq^2)&q^0\mp q^3\cr}} \cr
\then && u(\mini q)=
\mini{{1\over\sqrt{2q^0(q^0+q^3)}}}
{ \pmatrix{q^0+q^3&-q^1+iq^2\cr q^1+iq^2&q^0+q^3\cr}}
\rin SU(2)/U(1),~~q^2=0
\end{eqnarray}

$u(\mini q)$ with $q^2=0$ is constructed with the two  parameters
$\mini{{\rvec q\over q_0}}$  for the compact
manifold $SO(3)/SO(2)$ (2-sphere).
The three additional
'noncompact' parameters  of the real 5-di\-men\-sio\-nal Witt manifold
$SO^+(1,3)/SO(2)$
are contained in $s(\mini{p,\ul M})$ used above.

The fundamental Sylvester-Witt transmutators
are unitary
$u(\mini q)^\star=$ $u(\mini q)^{-1}$,
not hermitian.
For half-integer Lorentz representations, e.g. Weyl spinor fields
(chapter 2.),
the projections to the lightspaces $\L_+$ or $\L_-$ define the hermitian
Sylvester-Witt  projectors $p_\pm(q)$ as used in chapter 3.
\begin{eqnarray}
u(\mini q)\o{\bl 1_2+\si^3\over2}\o
u^\star(\mini q) &=& p_+(q)\o\bl 1_2\o p_+(q)
={\rho_{\ul j}q^{\ul j}\over2 q_0}\cr
u(\mini  q)\o{\bl 1_2-\si^3\over2}\o
u^\star(\mini q) &=&
p_-(q)\o\bl 1_2\o p_-(q)=
{\d\rho_{\ul j}q^{\ul j}\over2 q_0}
\end{eqnarray}

The integer spin Sylvester-Witt-transmutators are hermitian.
They start from the vector representation of the coset $SU(2)/U(1)$
\begin{eqnarray}
&&O(\mini q)=D^{({1\over2}|{1\over2})}(u(\mini q))
={ \pmatrix{
1&0&0&0\cr
0&1-{(q^1)^2\over q^0(q^0+q^3)}&-{q^1q^2\over q^0(q^0+q^3)}&{q^1\over q^0}\cr
0&-{q^1q^2\over q^0(q^0+q_3)}&1-{(q^2)^2\over q^0(q^0+q^3)}&{q^2\over q^0}\cr
0&-{q^1\over q^0}&-{q^2\over q^0}&{q^3\over q^0}\cr}}\rin SO(3)/SO(2)\cr
&&O(\mini q)
\mini{{ \pmatrix{{1\over\sqrt2}\cr0\cr0\cr\pm {1\over\sqrt2}}}}=
\mini{{1\over \sqrt2q^0}}
{ \pmatrix{q^0\cr\pm\rvec q\cr} },~~q^2=0,~~q\ne0
\end{eqnarray}

Instead of using a Sylvester basis with diagonal Lorentz matrix
$\eta$, it is more convenient for a Witt decomposition to
consider  a  Witt basis  with Lorentz matrix
$\io$
\begin{equation}
\hbox{Sylvester: }-\eta\cong{ \pmatrix{
-1&0&0&0\cr
0&1&0&0\cr
0&0&1&0\cr
0&0&0&1\cr}},~\hbox{Witt: }
-\io\cong{ \pmatrix{
0&0&0&1\cr
0&1&0&0\cr
0&0&1&0\cr
1&0&0&0\cr}}\end{equation}
The two bases are transformed into each other by $w$
\begin{eqnarray}
&& \mini{{1\over\sqrt2}{ \pmatrix{1\cr0\cr0\cr1\cr}}=
w{ \pmatrix{1\cr0\cr0\cr0\cr}},~~
{1\over\sqrt2}{ \pmatrix{1\cr0\cr0\cr-1\cr}}=
w{ \pmatrix{0\cr0\cr0\cr 1\cr}},~~~
w={ \pmatrix{
{1\over\sqrt2}&0&0&-{1\over\sqrt2}\cr
0&1&0&0\cr                   0&0&1&0\cr
{1\over\sqrt2}&0&0&{1\over\sqrt2}\cr}}    }
\end{eqnarray}

To distinguish the $SO(2)$-regime from the $SO(3)$-
and the $SO^+(1,3)$-regime
doubled underlined and Greek indices are used
\begin{equation}
 \hbox{vector indices for the $SO(2)$-Witt regime in a }
 \left\{ \begin{array}{l}
  \hbox{Sylvester basis: }\uul j=(0;1,2;3) \\
  \hbox{Witt basis: }\al=(0;1,2;3)
 \end{array} \right.
\end{equation}

The 2-parametric Sylvester-Witt transmutator $H(\mini q)$
\begin{eqnarray}
&&H(\mini q)^{\ul k} _{\al} =O(\mini q)_{\uul j}^{\ul k}\o w^{\uul j}_{\al}
\cong\mini{{1\over q^0}}{ \pmatrix{
\s{q^0\over\sqrt2}&\s0&\s0&\s-{q^0\over\sqrt2}\cr
\s{q^1\over\sqrt2}
&\s q^0-{(q^1)^2\over q^0+q^3}&\s{q^1q^2\over q^0+q^3}
&\s{q^1\over\sqrt2}  \cr
\s {q^2\over\sqrt2}&\s-{q^1q^2\over q^0+q^3}&\s q^0-{(q^2)^2\over q^0+q^3}
&{q^2\over\sqrt2}\cr
\s{q^3\over\sqrt2}
&\s - q^1&\s - q^2&\s {q^3\over\sqrt2}\cr}}\rin SO(3)/SO(2)
\end{eqnarray}
relates the bilinear forms for Sylvester and
Witt bases
\begin{eqnarray}
&&H(\mini q)_{\al}^{\ul k} \iota^{\al\be}H(\mini q)^{\ul j}_{\be}
=\eta^{\ul k\ul j},~~
 H(\mini q)_{\al}^{\ul k}\eta_{\ul k\ul j}H(\mini q)^{\ul j}_{\be}
 =\io_{\al\be}\cr
&&H(\mini q)^{\ul k}_0=\mini{{1\over\sqrt2q^0}}q^{\ul k}
\end{eqnarray}

\subsection{Harmonic Analysis of Massless Vector Fields}

The time representations in  the commutator
of the massless vector field are transformed from the Lorentz to the
Sylvester regime by a transmutator $\La(\mini{p,\ul M})$
\begin{eqnarray}
\com{\bl A^k}{\bl A^j}(x)
&&=\La(\mini{p,\ul M})^k_{\ul k}~
\mini{\int {d^3 \ul qe^{-i\ul{\rvec x}\rvec {\ul q}}\over(2\pi)^3\ul q_0} }
\brack{\bl A\bl A}^{\ul k\ul j}(\ul x_0,\ul q)~\La(\mini{p,\ul M})^j_{\ul j}
\end{eqnarray}
The transformation
$\La(\mini{p,\ul M})(x)=  \ul x=(\ul x_0,\ul{\rvec x})$ and
$\La(\mini{p,\ul M})(q)=
\ul q=(\ul q_0,\ul{\rvec q})$ allows the use of separate
time-energy and space-momentum coordinates.
This rest system, especially the mass scale $\ul M$,
has to be introduced by an  additional structure
$p^2=\ul M^2>0$. In contrast to the massive case,
it is not determined  by the momentum orbit of the massless field.

The commutator in the rest frame is given as follows
\begin{equation}
\mini{\brack{\bl A\bl A}^{\ul k\ul j}(\ul x_0,\ul q)=}
\mini{-\mu^2 \eta^{\ul k\ul j}i\sin \ul x_0\ul q_0
-{\mu^2 +\ep\si^2 \over2}{ \pmatrix{
\s i(\ul x_0\ul q_0\cos \ul x_0\ul q_0+\sin \ul x_0\ul q_0)
&\s \ul x_0\ul q^a\sin \ul x_0\ul q_0\cr
\s \ul x_0\ul q^b\sin\ul x_0\ul q_0&
\s{\ul q^a\ul q^b\over \ul q_0^2}
i(\ul x_0\ul q_0\cos \ul x_0\ul q_0-\sin \ul x_0\ul q_0) \cr}}
}
\end{equation}

The rotation $O(\ul q)$ transforms from the $SO(3)$ to the
$SO(2)$   regime, resulting in $\ul q^a\mape \de^a_3\ul q^0$.
Then one introduces by $w$ a Witt basis, $H(\ul q)=O(\ul q)\o w$
\begin{equation}
\brack{\bl A\bl A}^{\ul k\ul j}(\ul x_0,\ul q)=
H(\ul q)_\al^{\ul k}
{}~\brack{\bl A\bl A}^{\al\be}(\ul x_0) ~H(\ul q)_\be^{\ul j}
\end{equation}

In the Witt-regime
the 1st and 2nd component have
the commutators
\begin{equation}
\al,\be\in\{1,2\}:\brack{\bl A\bl A}^{\al\be}(\ul x_0)\cong
\mu^2 i\sin \ul x_0\ul q_0{ \pmatrix{
1&0\cr
0&1\cr}}\end{equation}
They have nontrivial circularity (polarization) $SO(2)\cong
U(1)$.
They carry two compact $U(1)$ time representations of energy $\ul q_0$
with 'positions' $\bl a^{1,2}$ analogous to the harmonic oscillator
\begin{equation}
 \al,\be\in\{1,2\}:
 \left\{ \begin{array}{l}
  \brack{\bl A\bl A}^{\al\be}(\ul x_0-\ul y_0)\de(\rvec {\ul q}-\rvec{\ul p})=
   \com{\bl a^{\al}(\rvec{\ul p},\ul y_0)}{\bl a^{\be}(\rvec {\ul q},\ul x_0)}
    \\
  \bl a^{\al}(\rvec {\ul q},\ul x_0) =\mu
   {\ro U^{\al}(\rvec {\ul q},\ul x_0) +\ro U^{\al \star}(\rvec{\ul q},
   \ul x_0)\over\sqrt2}
   =\bl a^{\al\star}(\rvec {\ul q},\ul x_0) \\
  \ro U^{\al}(\rvec {\ul q},\ul x_0)
   =\ro U^{\al}(\rvec{\ul q}) e^{i\ul x_0\ul q_0} \\
  \com{\ro U^{\al\star}(\rvec{\ul p})}{\ro U^{\be}(\rvec {\ul q})}
   =\de^{\al\be}\de(\rvec {\ul q}-\rvec{\ul p})
 \end{array} \right.
\end{equation}
and, therefore, have a particle interpretation (polarized photons)
with a Fock state
\begin{equation}
\al,\be\in\{1,2\}: \left\{ \begin{array}{l}
\Ffo{\brace{\bl A\bl A}^{\al\be}}(\ul x_0)\cong
\mu^2 \cos \ul x_0\ul q_0
{ \pmatrix{
1&0\cr  0&1\cr}}\\
\Ffo{\acom{\ro U^{\al\star}(\rvec{\ul p})}{\ro U^{\be}(\rvec {\ul q})} }
=\de^{\al\be}\de(\rvec {\ul q}-\rvec{\ul p}) \end{array} \right.
\end{equation}
The normalization  of the $SO(2)$-invariant is given by the square
of the massless photons
\begin{equation}
\al,\be\in\{1,2\}:~~
\Ffo{\brace{\bl A\bl A}^{\al\be}}(0)=\mu^2 \de^{\al\be}
\end{equation}

The $SO(2)$ trivial unpolarized contributions (0th and  3rd component)
\begin{equation}\al,\be\in\{0,3\}:~~\left\{ \begin{array}{l}
\brack{\bl A\bl A}^{\al\be}(\ul x_0)\cong
{\mu^2\over2}   { \pmatrix{
{i\ul x_0\ul q_0\over M_0}e^{-i\ul x_0\ul q_0} &N_0 i\sin \ul x_0\ul q_0\cr
N_0 i\sin \ul x_0\ul q_0 &{i\ul x_0\ul q_0\over M_0}e^{i\ul x_0\ul q_0}\cr}}
\\
{1\over M_0}=-\mini{{\mu^2+\ep\si^2\over\mu^2}},~~
N_0=\mini{{3\mu^2 +\ep\si^2 \over \mu^2}} \end{array} \right.
\end{equation}
carry one indefinite $U(1,1)$ time representation
as discussed in the quantum mechanical model (section 4.2.)
\begin{equation}
\al,\be\in\{0,3\}:~~\left\{ \begin{array}{l}
\brack{\bl A\bl A}^{\al\be}(\ul x_0) \de(\rvec {\ul q})=
{ \pmatrix{
\com{\bl a^3}{\bl a^0}& \com{\bl a^0}{\bl a^0}\cr
\com{\bl a^3}{\bl a^3}& \com{\bl a^0}{\bl a^3}\cr}}(\rvec {\ul q},\ul x_0) \\
\bl a^0(\rvec {\ul q},\ul x_0)=\mu
{\ro B(\rvec {\ul q},\ul x_0)+N_0\ro G^\x(\rvec q,\ul x_0)\over\sqrt2}
=\bl a^{3\x}(\rvec {\ul q},\ul x_0)\\
{ \pmatrix{
\ro B(\rvec {\ul q},\ul x_0)\cr  \ro G(\rvec {\ul q},\ul x_0)\cr}}
=e^{i\ul x_0\ul q_0}{ \pmatrix{
1&{i\ul x_0\ul q_0\over M_0}\cr 0&1\cr}}
{ \pmatrix{\ro B(\rvec {\ul q})\cr\ro G(\rvec{\ul q})\cr}}\\
\com{\ro B^\x(\rvec{\ul p})}{\ro G(\rvec {\ul q})}
=\com{\ro G^\x(\rvec{\ul p})}{\ro B(\rvec {\ul q})}
=\de(\rvec{\ul q}-\rvec{\ul p}) \end{array} \right.
\end{equation}

Therewith the harmonic analysis\footnote{
The complete
harmonic analysis of a massless Weyl field (chapter 3.)
has to include the Lorentz-Sylvester transmutator
\begin{equation}
\bl l^A(x)
=\mini{\int {d^3\ul q\over(2\pi)^{3/2}}}
p_+(p,\ul M,\ul q)_C^ A
\brack{e^{i\ul x \ul q}\ro U^C(\rvec {\ul q})
+e^{-i\ul x\ul q}\ro A^{\star C}(\rvec {\ul q})} ,~~
p_+(p,\ul M,\ul q)=s(p,\ul M)\o p_+(\ul q)\end{equation}
}
of the massless vector field exhibits
both definite unitary  $U(1)$ and indefinite unitary
$U(1,1)$ time representations
in the $(1,2)$ components (photons) and $(0,3)$ components
(Coulomb and 'gauge' degree of freedom) resp.
\begin{eqnarray}
&&\bl A^k(x)\mini{=\La(p,\ul M)^k_{\ul k}
\int{d^3\ul q\over\sqrt{(2\pi)^3 \ul q_0}}H(\ul q)^{\ul k}_{\al} }
\bl a^{\al}(\ul q,\ul x)\cr
&&\bl a^{\al}(\ul  q,\ul x)=\mu\pmatrix{
{\brack{\ro B(\rvec {\ul q})
+{i\ul x_0\ul q_0\over M_0}\ro G(\rvec{\ul q})}e^{i\ul x\ul q}
+N_0\ro G^\x(\rvec {\ul q})e^{-i\ul x\ul q}\over\sqrt2}\cr
{\ro U^1(\rvec {\ul q})e^{i\ul x\ul q}
+\ro U^{1\star}(\rvec {\ul q})e^{-i\ul x\ul q}\over\sqrt2}\cr
{\ro U^2(\rvec {\ul q})e^{i\ul x\ul q}+\ro U^{2\star}(\rvec {\ul q})
e^{-i\ul x\ul q}\over\sqrt2}\cr
{\brack{\ro B^\x(\rvec {\ul q})
-{i\ul x_0\ul q_0\over M_0}\ro G^\x(\rvec {\ul q})}e^{-i\ul x\ul q}
+N_0\ro G(\rvec {\ul q})e^{i\ul x\ul  q}\over\sqrt2}\cr}
\end{eqnarray}

\section{Masses and Coupling Constants}

As we have seen in the former chapters,
the masses and coupling constants, connected with
vector fields \`a la Syl\-ves\-ter and Witt
and their harmonic analysis, can quantify
characteristic invariants of the symmetries involved.

To have an experimentally relevant illustration, the
masses and coupling constants of the  vector
fields in the standard model\cite{Wei67} are considered.
The $Z$-boson coupling constant $g_Z$ and the electromagnetic
coupling constant $g_e$ with $g_e^2\simeq\mini{{4\pi\over 137}}$
measure the hypotenusis and height resp. of the rectangular
electroweak triangle   with the Weinberg angle $\th_w$.
This triangle\cite{Sal92b}\cite{Sal93}\cite{Sal94a}
gives the vector boson masses
in units of the symmetry breakdown Fermi mass $\ul M\simeq 123
\mini{{\ro{GeV}\over c^2}}$
\begin{eqnarray}
 \begin{array}{l}
 (m_Y,m_W,m_Z|m_e)=(g_Y,g_W,g_Z|g_e)\ul M\\
 ~~~\simeq (43.4,80.2,91.2|38.2)
  \mini{{\ro{GeV}\over c^2}} \end{array} ,~~
\left\{ \begin{array}{l}
g_Y^2+ g_W^2=g_Z^2\\
g_Yg_W=g_Zg_e \\
\tan\th_w={g_Y\over g_W} \end{array} \right.
\end{eqnarray}
Only $m_Z$ and $m_W$ are masses of {\it particles}.

A time representation as compact unitary group $U(1)$
\begin{equation}
\ro u(t)=e^{i\om t}\ro u,~~e^{i\om t}\in U(1)\end{equation}
comes with  a frequency $\om$  as $U(1)$ invariant scale.
The $U(1)$ scalar product defines the probability inducing
Fock form in a quantum theory
\begin{equation}
\sprod{\ro u}{\ro u}=\Ffo{\ro u^\star\ro u}=\expv 0{\ro u^\star\ro u}0=1
\end{equation}
For the {\it hermitian}
position and momentum
combinations in a Bose quantum theory,
the $U(1)\cong SO(2)$ time development symmetry leads to the
invariance of the $2\x 2$-matrix
\begin{equation}
{ \pmatrix{
\expv 0{\acom{\bl x}{\bl x}}0&
\expv 0{\acom{\bl p}{\bl x}}0\cr
\expv 0{\acom{\bl x}{\bl p}}0&
\expv 0{\acom{\bl p}{\bl p}}0\cr}}
={ \pmatrix{
{\om\over k}&0\cr 0& {k\over\om}\cr}}
\end{equation}

For the massive  vector fields (Sylvester), e.g. the $Z$-boson,
the mass  $m$ enters the time representations via
$\mini{\ul q_0=\sqrt{m^2+\rvec {\ul q}^2}}$.
The $SO(2)\cong U(1)$ time development normalization  is given
by the analogue matrix  replacing $(\bl x,\bl p)$ by
$(\bl Z^a,\bl G_{0a})$
\begin{equation} a,b\in\{1,2,3\}:~~
{ \pmatrix{
m\la\de^{ab}&0\cr 0&{1\over m\la}\de_{ab}\cr}}
\end{equation}
Here $\de^{ab},\de_{ab}$ take into account the three components of the
spin $SO(3)$ representation.  $\mini{\sqrt{m\la}}$ is the analogue to
$\mini{{\om\over k}}$ in the mechanical model.
The Lorentz-Sylvester
transmutator $\La(\mini{q,m})\rin SO^+(1,3)/SO(3)$
is constructed only with $m$ - not with $\la$.
It embeds  the Casimir invariant
$m^2\de^{ab}$ of $SO(3)$ in its Lorentz orbit
\begin{equation}\La(\mini{q,m})_a^k m^2\de^{ab}
\La(\mini{q,m})=-m^2\eta^{kj}+q^kq^j\end{equation}
Since the symmetry $SO(3)$ has only one independent invariant,
the  component
$m\la\de^{ab}$ in the invariant time development matrix above
is identified with $m^2\de^{ab}$.

{\it For the standard model, the $Z$-coupling constant
$m^2={m_Z^2\over\ul M^2}=g_Z^2$ turns out to be
the invariant spin $SO(3)\cong SU(2)/\{\pm\bl 1_2\}$ normalization
in the Lorentz group $SO^+(1,3)$.}

The indefinte unitary $U(1,1)$ time development
in the massless electromagnetic field
does not introduce additional masses
or coupling constants: Such a   time representation
\begin{equation}
{ \pmatrix{
\ro b(t)\cr  \ro g(t)\cr}}
=e^{i\om t}{ \pmatrix{1&{it\over M_0}\cr 0&1\cr}}
{ \pmatrix{
\ro b\cr  \ro g\cr}},~~
e^{i\om t}{ \pmatrix{1&{it\over M_0}\cr 0&1\cr}}\in U(1,1)\end{equation}
contains  two parameters, a frequency $\om$ as the invariant trace of the
time translation generator (up to $i$ the Hamiltonian matrix)
\begin{equation}
\mini{{1\over2}}
\tr{ \pmatrix{\om&{1\over M_0}\cr 0&\om\cr}} =\om
\end{equation}
and a mass $M_0$ in the nilpotent traceless contribution.
The mass $M_0$, however, reflects  only the choice of
a basis in the complex 2-dimensional time representation space and
- as a basis dependent quantity - is physically irrelevant.

For massless gauge vector fields the frequencies
$\om=\sqrt{\rvec {\ul q}^2}$ in the harmonic analysis are not
Lorentz invariant.
The gauge fixing sector
reflects the choice of a basis with the physically irrelevant
nontrivial gauge fixing parameter $\ep\si^2$
in the combination
\begin{equation}
\mini{{1\over M_0}=-{\mu^2+\ep\si^2\over \mu^2}}
\end{equation}

Unlike in the particle case with its definite $U(1)$ Fock state,
the indefinite $U(1,1)$ sesquilinear form of
the noncompact time representation
\begin{equation}
{ \pmatrix{
\sprod{\ro b}{\ro b}& \sprod{\ro b}{\ro g}\cr
\sprod{\ro g}{\ro b}& \sprod{\ro g}{\ro g}\cr}}=
{ \pmatrix{0&1\cr1&0\cr}}
\end{equation}
does not contribute  to the probabilities
in a quantum theory.
Especially for nonabelian quantum gauge vector fields,
e.g. the Fadeev-Popov fields\cite{Kug78}\cite{Nak90}
take care of  the metrical
structure
\footnote{
Also the fermionic Fadeev-Popov fields have a harmonic analysis
with $U(1,1)$ time representations\cite{Sal94b}.}
of $U(1,1)$.

For the massless electromagnetic field,
the $U(1)$ time representations
for the photons arise with frequencies
$\mini{\ul q_0=\sqrt{\rvec {\ul q}^2}}$. The $U(1)\cong SO(2)$
time development normalization is
given by
\begin{equation}
\al,\be\in\{1,2\}:~~
{ \pmatrix{
\mu^2\de^{\al\be}&0\cr 0&{1\over\mu^2}\de_{\al\be}\cr}}
\end{equation}
Here $\de^{\al\be},\de_{\al\be}$ takes into account the two components
of the polarization $SO(2)$ representation.  The Sylvester-Witt
transmutator $O(\mini q)\rin SO(3)/SO(2)$
embeds the $SO(2)$ invariant
$\mu^2\de^{\al\be}$ in its $SO(3)$-orbit,
followed by an embedding with the Lorentz-Sylvester transmutator
$\La(p,\ul M)$ in its Lorentz $SO^+(1,3)$-orbit.

{\it For the standard model,
the electromagnetic coupling constant
$\mu^2={m_e^2\over\ul M^2}=g_e^2$ turns out to be
the  circularity $SO(2)\cong U(1)$
normalization in the Lorentz group $SO^+(1,3)$.}

The three mass parameters involved - or one mass parameter
and  two mass ratios - can be related to the
Lorentz-Sylvester-Witt
chain of the three associated groups
\begin{equation}
SO^+(1,3)\supnoteq SO(3)\supnoteq SO(2):~~\left\{ \begin{array}{l}
\ul M,~~m_Z,~~m_e\\
\ul M,~~\sin 2\th_w=2{m_e\over m_Z},~~\al_e={m_e^2\over4\pi\ul M^2}
\end{array} \right.
\end{equation}

\end{document}